# 1. Introduction

A digital approach to geometry and topology plays an important role in analyzing n-dimensional digitized images arising in computer graphics as well as in many areas of science including neuroscience, medical imaging, industrial inspection, geoscience and fluid dynamics. Concepts and results of the digital approach are used to specify and justify some important low-level image processing algorithms, including algorithms for thinning, boundary extraction, object counting, and contour filling.

Usually, a digital object is equipped with a graph structure based on the local adjacency relations of digital points [8]. In papers [9-10], a digital n-surface was defined as a simple undirected graph and basic properties of n-surfaces were studied. Paper [9] analyzes a local structure of the digital space $Z^n$. It is shown that $Z^n$ is an n-surface for all n>0. In paper [10], it is proven that if A and B are n-surfaces and A⊆B, then A=B.

X. Daragon et al. [6-7] studied partially ordered sets in connection with the notion of n-surfaces. In particular, it was proved that (in the framework of simplicial complexes) any n-surface is an n-pseudomanifold, and that any n-dimensional combinatorial manifold is an n-surface. In paper [23], M. Smyth et al. defined dimension at a vertex of a graph as basic dimension, and the dimension of a graph as the sup over its vertices. They proved that dimension of a strong product G × H is dim ( G ) + dim ( H ) (for non-empty graphs G and H).

An interesting method using cubical images with direct adjacency for determining such topological invariants as genus and the Betti numbers was designed and studied by L. Chen et al. [3].

E. Melin [22] studies the join operator, which combines two digital spaces into a new space. Under the natural assumption of local finiteness, he shows that spaces can be uniquely decomposed as a join of indecomposable spaces. In papers [2,14], digital covering spaces were classified by using the conjugacy class corresponding to a digital covering space.

A digital n-manifold, which we regard in this paper, is a special case of a digital n-surface.
We define the complexity of digital n-manifolds similar to the notion of complexity of continuous 3- and 4-manifolds studied in [20-21]. We introduce compressed digital n-manifolds and show that any n-manifold can be transformed to a compressed one by transformations retaining the connectedness and the dimension of the given n-manifold. We show that a digital n-manifolds are classified by complexity and the compressed n-manifolds which are equivalent to the given ones.

## 2. Digital spaces, contractible graphs and contractible transformations

A digital space G is a simple undirected graph G=(V,W), where V={$v_1,v_2,...v_n,...$} is a finite or countable set of points, and W = {$(v_p v_q)$,....}⊆V×V is a set of edges provided that $(v_p v_q)=(v_q v_p)$ and $(v_p v_p)\notin W$. Such notions as the connectedness, the adjacency, the dimensionality and the distance on a graph G are completely defined by sets V and W. Further on, if we consider a graph together with the natural topology on it, we will use the phrase 'digital space". We use the notations $v_p\in G$ and $(v_p v_q)\in G$ if $v_p\in V$ and $(v_p v_q)\in W$ respectively if no confusion can result. |G| denotes the number of points in G.
Since in this paper we use only subgraphs induced by a set of points, we use the word subgraph for an induced subgraph. We write H⊆G. Let G be a graph and H⊆G. G-H will denote a subgraph of G obtained from G by deleting all points belonging to H. For two graphs G=(X,U) and H=(Y,W) with disjoint point sets X and Y, their join G⊕H is the graph that contains G, H and edges joining every point in G with every point in H. Points $v_p$ and $v_q$ are called adjacent if $(v_p v_q)\in W$. The subgraph O(v)⊆G containing all points adjacent to v (without v) is called the rim or the neighborhood of point v in G, the subgraph U(v)=v⊕O(v) is called the ball of v. Graphs can be transformed from one into another in a variety of ways. Contractible transformations of graphs seem to play the same role in this approach as a homotopy in algebraic topology [17-18].

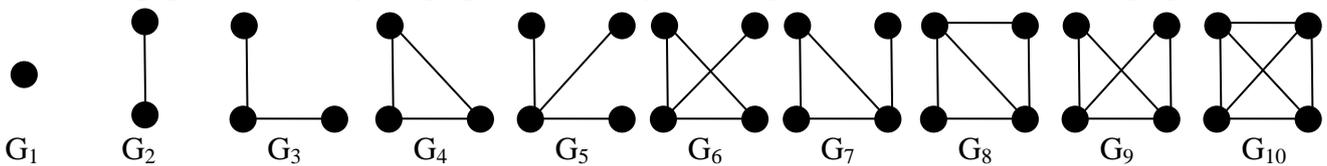

Figure 1. Contractible graphs with the number of points n<5.

A graph G is called contractible (fig. 1), if it can be converted to the trivial graph by sequential deleting simple points. A point v of a graph G is said to be simple if its rim O(v) is a contractible graph.

An edge (vu) of a graph G is said to be simple if the joint rim O(vu)=O(v)∩O(u) is a contractible graph. In [17], it was shown that if (vu) is a simple edge of a contractible graph G, then G-(vu) is a contractible graph. Thus, a contractible graph can be converted to a point by sequential deleting simple points and edges. In fig.1, $G_{10}$ can be converted to $G_9$ or $G_8$ by deleting a simple edge. $G_9$ can be converted to $G_7$ or $G_6$ by deleting a simple edge. $G_6$ can be converted to $G_5$ by deleting a simple edge. $G_7$ can be converted to $G_4$ by deleting a simple point. $G_5$ can be converted to $G_3$ by deleting a simple point. $G_3$ can be converted to $G_2$ by deleting a simple point. $G_2$ can be converted to $G_1$ by deleting a simple point.

Deletions and attachments of simple points and edges are called contractible transformations. Graphs G and H are called homotopy equivalent or homotopic if one of them can be converted to the other one by a sequence of contractible transformations.

Homotopy is an equivalence relation among graphs. Contractible transformations retain the Euler characteristic and homology groups of a graph [18].

Properties of graphs that we will need in this paper were studied in [12,17-18].

**Proposition 2.1** • Let G be a graph and v be a point (v∉G). Then v⊕G is a contractible graph. If K is a clique then K⊕G is a contractible graph.
- Let G be a contractible graph and S(a,b) be a disconnected graph with just two points a and b. Then S(a,b)⊕G is a contractible graph.
- Let G be a contractible graph with the cardinality |G|>1. Then it has at least two simple points.
- Let H be a contractible subgraph of a contractible graph G. Then G can be transformed into H sequential deleting simple points.
- Let graphs G and H be homotopic. G is connected if and only if H is connected. Any contractible graph is connected.

For any terminology used but not defined here, see Harary [15].

## 3. Digital n-dimensional manifolds

There is an abundant literature devoted to the study of different approaches to digital lines surfaces and spaces used by researchers, just mention some of them [1, 16, 19]. A digital n-manifold is a special case of a digital n-surface defined and investigated in [9].

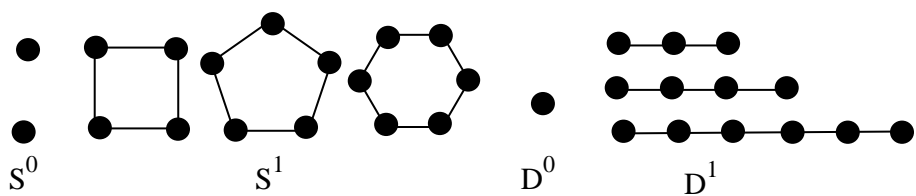

Figure 2. Zero- and one-dimensional spheres $S^0$ and $S^1$ and zero- and one-dimensional disks $D^0$ and $D^1$.

**Definition 3.1.** A *digital 0-dimensional sphere* is a disconnected graph $S^0$(a,b) with just two points a and b (fig. 2).

To define digital n-spheres, n>0, we will use a recursive definition. Suppose that we have defined digital k-spheres for dimensions 0≤k≤n-1.

**Definition 3.2.** ● A connected digital space M is called a *digital n-sphere, n>0,* if for any point v∈M, the rim O(v) is an (n-1)-sphere and the space M-v is a contractible graph (fig. 2-3) [10-11].
● Let M be a digital n-sphere, n>0, and v be a point belonging to M. The space D=M-v is called *a digital n-disk* (fig. 2-3).

Obviously, a digital n-disk D can be represented by the union D=∂D∪IntD: ∂D=O(v) is an (n-1)-sphere, IntD=D-∂D. The rim O(x) is an (n-1)-sphere if a point x∈IntD and the rim O(x) is an (n-1)-disk if x∈∂D (fig. 2-3). Spaces IntD and ∂D are called *the interior* and *the boundary* of D respectively. Further on, we say "space" to abbreviate "digital space", if no confusion can result.

**Definition 3.3.** ● A connected space M is called *an n-dimensional manifold,* n>1, if the rim O(v) of any point v is an (n-1)-dimensional sphere.
● Let M be an n-manifold and a point v belong to M. Then the space N=M-v is called an n-manifold with the (spherical) boundary ∂N=O(v) and the interior IntN=N-∂N.

Evidently, N=∂N∪IntN, ∂N=O(v) is an (n-1)-sphere, IntN=M-(v⊕O(v)). It is not hard to check that the rim O(x) is an (n-1)-sphere if a point x∈IntN and the rim O(x) is an (n-1)-disk if x∈∂N.
It follows from definitions 3.2 and 3.3, that if for some point v belonging to an n-manifold M, the space M-

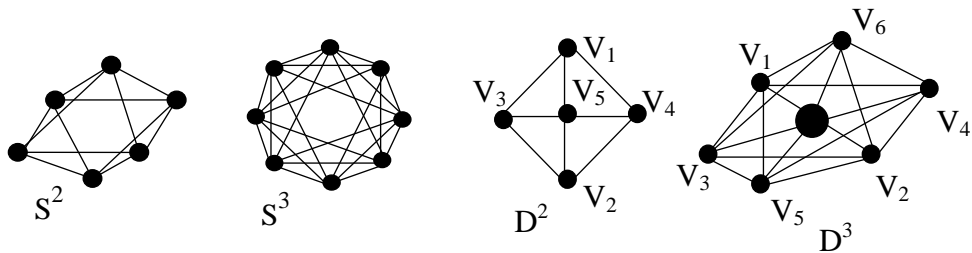

Figure 3. Minimal 2- and 3-dimensional spheres and disks.

v is not a contractible space, then M is not an n-sphere. Note that in this paper, we use only connected n-manifolds, n>0. For an n-surface studied in [9], the rim of a point is an (n-1)-surface whereas for an n-manifold the rim of any point is an (n-1)-sphere. A digital 2-dimensional torus T and a digital 2-dimensional projective plane P are depicted in fig. 4. As one can see, the rim of any point in T and in P is a digital 1-sphere.

**Proposition 3.1.** The join $S^n_{min}=S^0_1 \oplus S^0_2 \oplus \ldots S^0_{n+1}$ of (n+1) copies of the zero-dimensional sphere $S^0$ is an n-sphere.
**Proof.** The proof is by induction on the dimension n. For n=1, the proposition is plainly true. Assume that the proposition is valid whenever n<k. Let n=k and $v \in S^n_{min}$. Then $O(v)=S^{n-1}_{min}$ by construction, i.e., O(v) is the minimal (n-1)-sphere. Therefore, $S^n_{min}$ is an n-manifold. Evidently, $S^n_{min}-v = u \oplus O(v)$, where u is the only point, which does not belong to O(v). This is a contractible space according to proposition 2.1. Hence, $S^n_{min}$ is an n-sphere. □

**Definition 3.4.** The join $S^n_{min}=S^0_1 \oplus S^0_2 \oplus \ldots S^0_{n+1}$ of (n+1) copies of the zero-dimensional sphere $S^0$ is called *the digital minimal n-sphere* (fig. 2-3) [10].

**Proposition 3.2.** Any n-sphere M can be converted to the minimal n-sphere $S_{min}$ by contractible transformations.
**Proof.** Since $M-v_1=D=\partial D \cup IntD$ is a digital n-disk, i.e., a contractible space, we glue a simple point $x_1$ to M in such a manner that $O(x_1)=D$. In the obtained space $N=M\cup x_1$, any point y belonging to IntD is simple because $O(y)_N=x_1 \oplus (O(y)$, i.e., a contractible space according to proposition 2.1. Therefore, this point can be deleted from N. Delete all such points and consider the obtained space $M_1=S^0(v_1,x_1)\oplus O(v_1)$, where $O(v_1)=O(x_1)$. $M_1$ is a digital n-sphere by construction. Let a point $v_2 \in O(v_1)$. $M_1-v_2=D_2$ is a digital n-disk.

Using the same procedure as above, we obtain a digital-sphere $M_2 = S^0(v_1,x_1) \oplus S^0(v_2,x_2) \oplus O(v_1v_2)$. Acting in the same way, we finally obtain a minimal n-sphere $M = S^0(v_1,x_1) \oplus S^0(v_2,x_2)...\oplus S^0(v_{n+1},x_{n+1}) \oplus O(v_1v_2)$. □

**Proposition 3.3.** Let M be an n-sphere and G be a contractible space contained in M. Then the space M-G is a contractible space.
**Proof**. The proof is by induction on the dimension n. For n=1, the proposition is verified directly. Assume that the proposition is valid whenever n<k+1. Let n=k+1. Let M be an n-sphere and G be a contractible subspace of M.
Since G is contractible, there is a point x belonging to G and simple in G, i.e., $O(x) \cap G$ is contractible according to proposition 2.1. Since O(x) is an (n-1)-sphere, then by the induction hypothesis, O(x)- $G = O(x) \cap (M-G)$ is also contractible. Hence, x is simple in M-G. Therefore, $G_1 = G-x$ is a contractible space and $M-G_1 = (M-G) \cup x$ is homotopy equivalent to M-G. Acting in the same way we finally convert the space G to a point v and the space M-G to the space M-v. Spaces M-v and M-G are homotopy equivalent by construction. Since M-v is contractible then M-G is a contractible space. This completes the proof. □

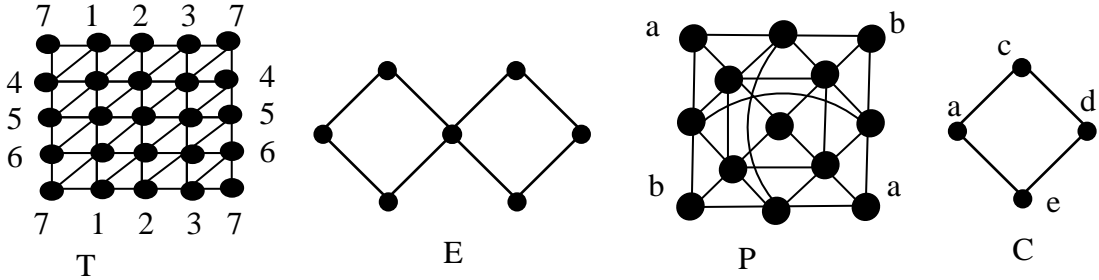

Figure 4. A digital 2-dimensional torus T and a digital 2-dimensional projective plane P. By sequential deleting simple points and edges, T-{1} can be converted to E and P-{b} can be converted to C.

**Proposition 3.4.** Let M be an n-manifold, G and H be contractible subspaces of M and v be a point in M. Then:
  (1) Subspaces M-G, M-H and M-v are all homotopy equivalent to each other.
  (2) M-G is a connected space.
**Proof**. To prove (1), notice that repeating word for word the proof of proposition 3.3, we show that M-G is homotopy equivalent to M-v, where v is any point belonging to G. Similarly, M-H is homotopy equivalent to M-u, where u is any point belonging to H. Consider a path P(v,u) connecting points v and u. Since P is a contractible space, then for the same reason as above, M-P, M-v and M-u are homotopy equivalent. Hence, M-G, M-H, M-v and M-u are all homotopy equivalent.
To prove (2), notice that M-v is a connected space by construction. As M-G and M-v are homotopy equivalent, then M-G is connected according to proposition 2.1. □

In a common sense, a digital n-dimensional sphere $S^n$ is the simplest n-manifold since it contains the smallest number of points compared to any other n-manifold [10].
Notice that there is a variety of ways to build new n-manifolds from given ones. As an example, consider two n-manifolds M and N. Suppose that there are points $v \in M$ and $u \in N$ such that O(v) and O(u) are isomorphic and f: O(v)→O(u) is the isomorphism of O(v) to O(u). It is not hard to check that the space W=(M-v)#(N-u) obtained by deleting points v and u from M and N and identifying each point in O(v)⊆M-v with its counterpart in O(u)⊆N-u is an n-manifold.

**4. R-Transformations of n-manifolds**

Our purpose in this section is to describe transformations of n-manifolds retaining the local topology of M, i.e., the topology of the neighborhood of a points. Paper [12] introduces and studies transformation of graphs involving cover graphs. The next construction is a special case of a cover graph. We take into account the topology and the dimension of an n-manifold. Consider a transformation increasing the number of points in a given digital n-manifold.

**Definition 4.1.** Let M be an n-manifold, v and u be adjacent points in M and (vu) be the edge in M. Glue a point x to M, where O(x)=v⊕u⊕O(vu), and delete the edge (vu) from the space. This pair of contractible transformations is called the replacement of an edge with a point or R-transformation, R: M→N. The obtained space N is denoted by N=RM=(M∪x)-(vu).

The following proposition is a direct consequence of theorem 4.2 proven in [11].

**Proposition 4.1.** Let M be an n-manifold and N=RM be a space obtained from M by an R-transformation. Then N is homotopy equivalent to M.
**Proof.** In definition 4.1, v⊕u⊕O(vu) is a contractible space according to proposition 2.1.. Therefore, x is a simple point, which is attached to M. In the obtained space P=M∪x, the edge (vu) is simple because O(vu)$_P$=x⊕O(vu), and can be deleted. Hence, N=rM=(M∪x)-(vu) is homotopy equivalent to M. □

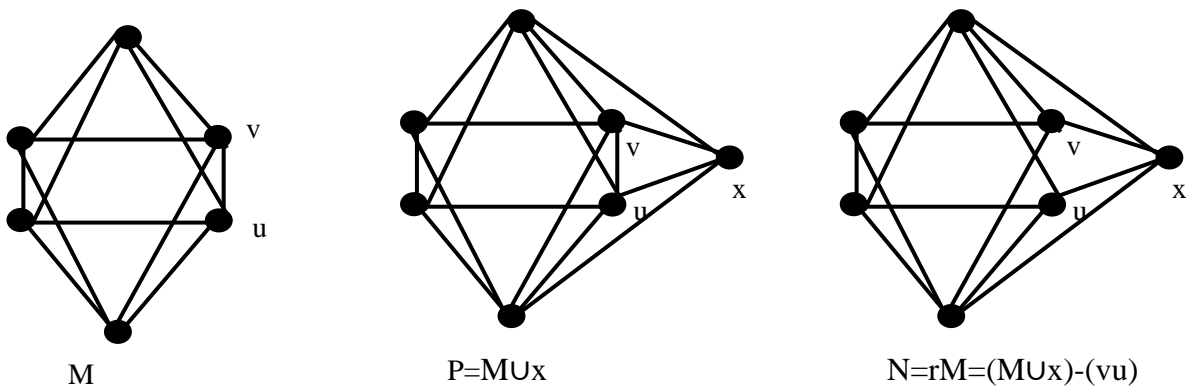

Figure 5. M is a digital 2-sphere with six points . P is a digital space obtained by attaching a simple point x. N=rM=(M∪x)-(vu) is a digital 2-sphere consisting of seven points.

The following proposition is a direct consequence of theorem 4.2 proven in [11].

**Proposition 4.2.** Let M be an n-manifold and N=RM be a space obtained from M by an R-transformation. Then N is a digital-manifold.

**Corollary 4.1.** Let M be an n-manifold and N=RM be a space obtained from M by an R-transformation. Then N is an n-manifold homotopy equivalent to M.

Figure 5 shows an R-transformation of a minimal 2-sphere M containing six points. P=M∪x is a space obtained by gluing a simple point x to M in such a way that O(x)=v⊕u⊕O(vu).The space N=RM is a digital 2-sphere consisting of seven points. In fact, an R-transformation is a digital homeomorphism because it retains the dimension and other local topological features of an n-manifold.
R-Transformations increase the number of points in a given n-manifold M retaining the global topology (the homotopy type of M) and the local topology (the homotopy type and the dimension of the neighborhood of any point).
By analogy with graph theory, call n-manifolds M and N *homeomorphic* if one of them can be transformed to the other one by a sequence of R-transformations.

**5. Compression of n-manifolds**

Our purpose now is to reduce the number of points and the number of edges in an n-manifold M by using contractible transformations, which retain the topology of M.
In graph theory, the contraction of points x and y in a graph G is the replacement of x and y with a point z such that z is adjacent to the points to which points x and y were adjacent. We use a similar definition for a digital n-disk.

**Definition 5.1.** Let $D=\partial D\cup IntD$ be a digital n-disk lying in a digital n-manifold M. The contraction of D is the replacement of all points belonging to IntD with a point z such that z is adjacent to all points belonging to $\partial D$.

For example, a digital n-sphere M can be converted to a minimal n-sphere by sequential contracting digital n-disks, as it follows from the proof of proposition 3.2.

**Definition 5.2.** A digital n-manifold M is called compressed if any digital n-disk lying in M is the ball of some point.

**Proposition 5.1.** Let $D=\partial D\cup IntD$ be a digital n-disk lying in a digital n-manifold M, and let

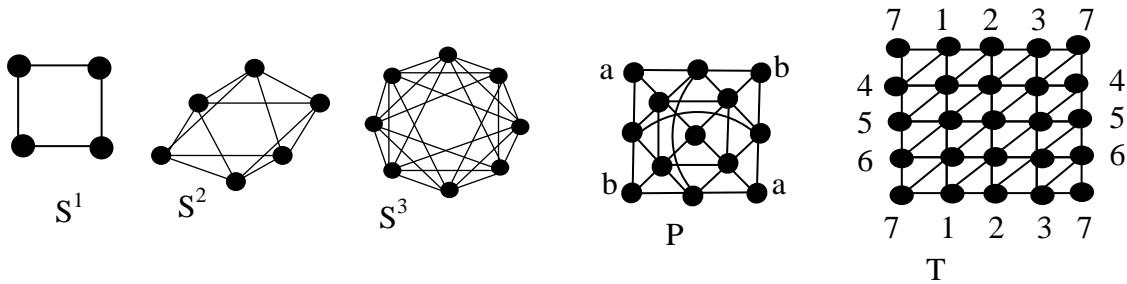

Figure 6. Compressed digital 1-, 2- and 3-spheres $S^1$, $S^2$ and $S^3$, a compressed digital 2-dimensional torus T, and a compressed digital 2-dimensional projective plane P.

$N=(M\cup z)-IntD$ be the space obtained by the contraction of D. Then N is a digital n-manifold homotopy equivalent to M.
**Proof.** Attach a point z to M in such a way that $O(z)=D$. Since D is a contractible space according to definition 3.2. Therefore, z is a simple pair, and the space $P=M\cup z$ is homotopy equivalent to M. It follows from construction of M and D, that for any point y belonging to IntD, the rim $O(y)_P$ is the cone $z\oplus(O(y))$, i.e. a contractible space according to proposition 2.1. Therefore, y is a simple point of P, and can be deleted. Delete all such points from P. The obtained space $N=(M\cup z)-IntD$ is homotopy equivalent to P and M.
The rim $O(z)$ of z in N is a digital (n-1)-sphere, $O(z)=\partial D$. The rim of any other point belonging to N is a digital (n-1)-sphere by construction of N. Therefore, N is a digital-manifold homotopy equivalent to M. The proof is complete. □

The following corollary is a direct consequence of proposition 5.1.

**Corollary 5.1.** A digital n-manifold M can be converted to a compressed form by sequential contracting digital n-disks.

Compressed 1-, 2- and 3-spheres are shown in fig. 6. Evidently, if M is an n-sphere, then CM is the minimal n-sphere $S^n_{min}$. The compressed 2-torus T contains sixteen points, the compressed two-dimensional projective plane P contains eleven points (see fig. 6).

**Proposition 5.2.** Let M be a compressed n-manifold and points v and u be adjacent in M. Then there is a minimal 1-sphere S consisting of four points, lying in M, and containing points v and u.

**Proof**. Consider The union U(v)∪U(u) of balls of adjacent points v and u. If U(v)∪U(u)=D=∂D∪IntD is a digital n-disk, then IntD={v,u} and ∂D=U(v)∪U(u)-{v,u}=O(v)∪O(u)-{v,u}  Since ∂D is a digital (n-1)-sphere according to definition 3.2, then any point x belonging to O(v)-O(u) is not adjacent to any point y belonging to O(y)-O(x) by construction of D (see fig. 7(a)). If U(v)∪U(u) is not a digital n-disk, then O(v)∪O(u)-{v,u} is not a digital (n-1)-sphere, i.e., there are adjacent points x and y such that x belongs to O(v)-O(u) , and y belongs to O(y)-O(x) (see fig. 7(b)), and {v,u,y,x} is a digital 1-sphere. Since M is a compressed digital n-manifold, then U(v)∪U(u) is not a digital n-disk, and there is a digital 1-sphere {v,u,y,x}. The proof is complete. □

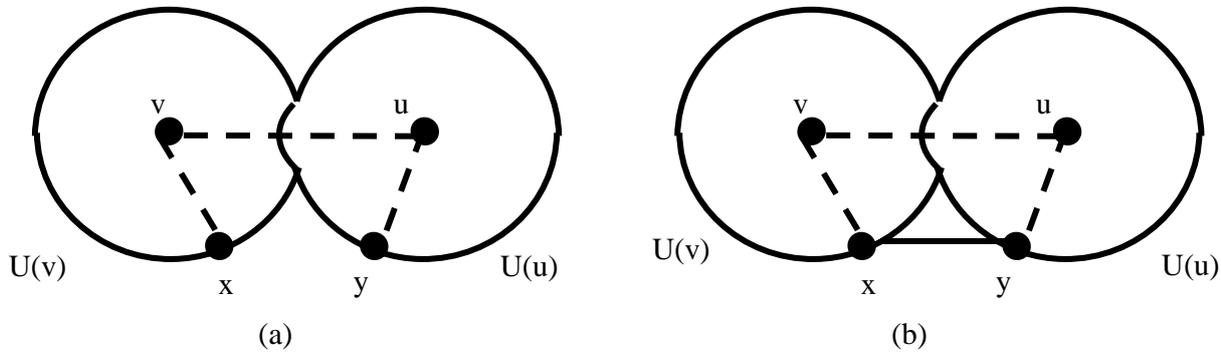

Figure 7. (a) U(v)∪U(u) is a digit l n-disk. (b) U(v)∪U(u) is not a digit l n-disk. {v,u,y,x} is a digital 1-sphere.

**Definition 5.3.** Let M be an n-manifold. Denote *CM* the compressed n-manifold obtained from M by sequential contracting digital n-disks. Call CM *the* compression of M.

**Proposition 5.3.** Let M be an n-manifold and CM be the compression of M. Then for any point v∈M and for any point u∈CM, spaces M-v and CM-u are homotopy equivalent.

Proposition 5.3 follows directly from proposition 3.4. In general, |M| is greater than or equal to |CM|. In this sense, CM is simpler than M and can serve as a representative of the family all n-manifolds homeomorphic to M.

## 6. Classification of digital n-manifolds

In topology, the classification of n-dimensional manifolds is an old problem. The classification was done long ago for n = 2, i.e. for closed surfaces. However, classifying the manifolds in dimension 3- and higher turn out to be much more complex problem [5].
 The notion of complexity of continuous smooth, closed and orientable 3- and 4-manifolds was defined and studied, for example in papers [4, 20-21]. A natural notion of complexity of n-dimensional manifold is the minimal number of highest dimensional simplexes in a triangulation of the manifold. We give the definition of complexity of a digital n-manifold based on a similar idea.

**Definition 6.1.** Let M be a digital n-manifold with the number of points |M| and let CM be the compression of M with the number of points |CM|. The complexity of M, denoted by com(M) is |CM|.

The following statement is checked directly.

**Proposition 6.1.** Let M be an n-manifold and |M| be the number of points in M. Then:
- If n=1, then com(M)=4.
- If n>1, then 2n+2≤com(M)≤|M|.

One can use various tools for the classification of digital n-manifolds [2, 14]. One path based on the complexity and homeomorphism is considered below. Usually, the "classification" means up to an appropriate equivalence, here it means to define a sequence of compressed n-manifolds $M_1, M_2,\ldots,$ any n-manifold M is homeomorphic to one of them.

Note first, that for any positive number N there is a finite number of n-manifolds with N points and the dimension $n \leq \frac{N-2}{2}$.

Now we can describe an algorithm for the classification of digital n-manifolds.
- For given n and N, $2n+2 \leq N$, find all compressed n-manifolds with s points, $2n+2 \leq s \leq N$. Denote B(n,N) the set of these n-manifolds. Obviously, B(n,N) contains a finite number of elements.
- Let M be an n-manifold with the number of points $|M|=N+1$. Convert M to its compression CM. If M=CM, then CM$\notin$B(n,N), i.e., M belongs to the set B(n,N+1). If $|M|>|CM|$, then M is homeomorphic to CM$\in$B(n,N).

As an example, table 1 shows all compressed manifolds for $N \leq 16$ and $n \leq 7$.

|     | N=2 | N=4 | N=6 | N=8 | N=10 | N=11 | N=12 | N=14 | N=16 |
|-----|-----|-----|-----|-----|------|------|------|------|------|
| n=0 | $S^0$ |     |     |     |      |      |      |      |      |
| n=1 |     | $S^1$ |     |     |      |      |      |      |      |
| n=2 |     |     | $S^2$ |     |      | P    |      |      | T    |
| n=3 |     |     |     | $S^3$ |      |      |      |      |      |
| n=4 |     |     |     |     | $S^4$ |      |      |      |      |
| n=5 |     |     |     |     |      |      | $S^5$ |      |      |
| n=6 |     |     |     |     |      |      |      | $S^6$ |      |
| n=7 |     |     |     |     |      |      |      |      | $S^7$ |

Table 1. Compressed n-manifolds for $n \leq 7$, $N \leq 16$.

Here $S^n$, n=0,…7, are compressed (minimal) n-spheres, P is the compressed 2-dimensional projective plane, T is the compressed 2-dimensional torus (see fig. 6). Obviously, Com($S^2$)=6, Com(P)=11, Com(T)=16. Thus, a digital 2-sphere is the simplest 2-manifold, because it contains the minimal number of points, a projective plane P with com(P)=11 is more complex than $S^2$, a torus T with com(T)=16 is more complex than P. As one can see from table 1, for some numbers N there is no compressed n-manifold at all. Compressed n-manifolds do not exist for N=1,3,5,7,9,13,15.

More detailed characterization of n-manifolds in general and compressed n-manifolds in particular is based on the structure of spaces, which are n-manifolds without a point (or without a contractible subspace). It follows from proposition 4.4, that if M and X are homeomorphic n-manifolds, then M-v and X-u are homotopy equivalent spaces and if M-v and X-u are not homotopy equivalent, then M and X are not homeomorphic.

As an illustration, consider n-manifolds without a point v in table 2.

| Space | $S^n$-v, (n-1,…7) | $P^2$-v | $T^2$-v |
|-------|-------------------|---------|---------|
| Homotopy type | 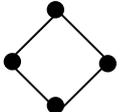 | 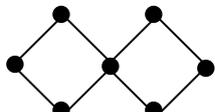 | |

Table 2. Homotopy type of 2-manifolds without a point.

Spheres $S^1,\ldots S^7$ are not homeomorphic, but $S^1$-v,… $S^7$-v, are homotopy equivalent to the one-point graph. It is easy to check directly that $T^2$-v is homotopy equivalent to the space E depicted in fig. 4, and $P^2$-v is homotopy equivalent to the space C (see fig.4). $T^2$-v and $P^2$-v are homotopy non-equivalent and $T^2$ and $P^2$ are not homeomorphic 2-manifolds.

Computationally it may be easier to analyze the structure of M-v than to find and analyze the compression CM of M. Since M-v is not a manifold, it can be transformed to its simplest form with the minimal number of points by a sequential deleting simple points and edges. One way to do this is the following:
1. Obtain from M-v the space $M_1$ by sequential gluing all simple edges.

2. Obtain from $M_1$ the space $M_2$ by sequential deleting all simple points.
3. Repeat steps 1-2 until we obtain the space C(M-v) with the minimal number of points. We say that C(M-v) is the compression of M-v.

Thus, one can use three elements for the classification of a digital n-manifold:

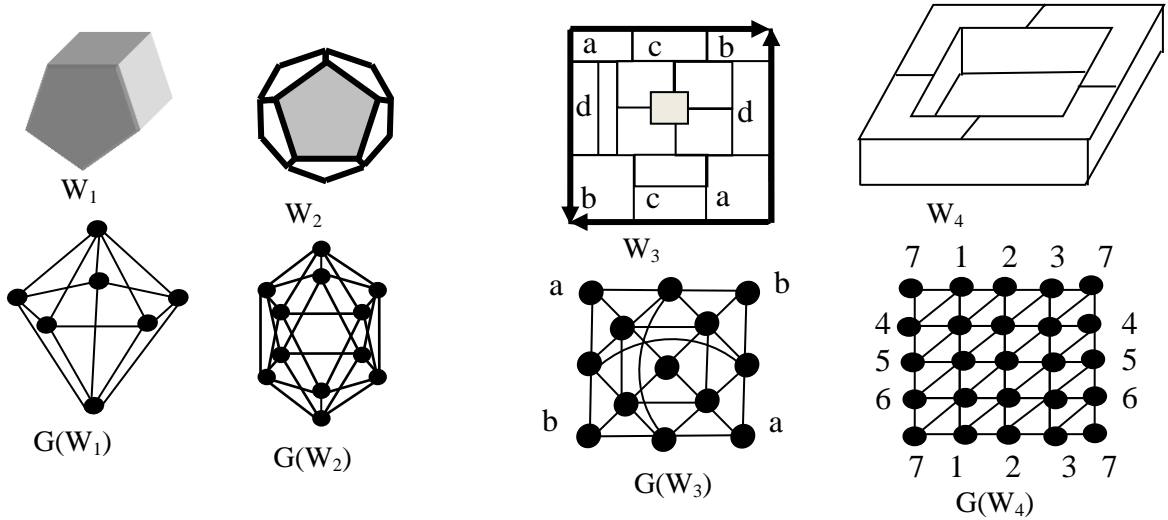

Figure 8. $W_1$ and $W_2$ are covers of a continuous 2-sphere $S^2$. Intersection graphs $G(W_1)$ and $G(W_2)$ are digital 2-spheres. $W_3$ and $W_4$ are covers of a projective plane P, and a torus T respectively. Intersection graphs $G(W_3)$ and $G(W_4)$ are digital 2-manifolds.

- The complexity com(M) of M (a positive integer).
- The compression C(M-v) of M-v (a digital space).
- The compression CM of M (a digital n-manifold).

It is clear, that the classification of digital n-manifolds in this context, is a computational problem, which is understood to be a task that is in principle capable to being solved by a computer (i.e. the problem can be stated by a set of mathematical instructions).

Finally, let us mention that the connections between continuous closed surfaces and digital 2-manifolds were studied in [10, 12-13]. In fig.8, $W_1$ and $W_2$ are covers of a continuous sphere S, $W_3$ and $W_4$ are covers of a projective plane P and a torus T. The intersection graphs $G(W_1)$, $G(W_2)$, $G(W_3)$ and $G(W_4)$ are digital 2-manifolds.

In {13}, LCL discretization schemes of the plane are defined and studied. In particular, it is shown that for any LCL discretization of the plane, the intersection graph of the discretization is necessarily a digital 2-manifold. Notice that integrating topological features into discretization schemes in order to generate topologically correct digital models of anatomical structures is critical for many clinical and research applications.

If we show that LCL discretization schemes can be applied to closed surfaces, we can use digital tools studied in this paper for classification of closed surfaces.

We end with a problem for further study. If we find out correct discretization schemes on continuous n-dimensional manifolds, we can construct digital models of these manifolds as the intersection graphs of discretization schemes. After this, we can consider the classification of digital n-manifolds as the classification of their continuous counterparts, and apply characteristics of a digital n-manifold to its continuous counterpart. For example, in this way we can classify continuous 3-dimensional manifolds.